\documentclass[screen,authorversion]{acmart}

\usepackage{cleveref,subcaption}
\usepackage{float}

\definecolorset{HTML}{}{}{%
  skills,dc3cbc;%
  explanations,8cb41e;%
  control,0046dc;%
  strategies,ff6e00;%
  theme,9900ff;%
  feature,f2f2f2%
}

\renewcommand{\quote}[1]{\textit{``#1''}}
\newcommand{\rem}{[...] }

\def\feature#1{\colorbox{gray!12}{\textbf{#1}}}

\usepackage{framed}
\usepackage{xcolor,colortbl}
\usepackage{acmart-taps}

\definecolor{shadecolor}{rgb}{0.89,0.74,0.99}%

\copyrightyear{2025}
\acmYear{2025}
\setcopyright{rightsretained}
\acmConference[LAK 2025]{LAK25: The 15th International Learning Analytics and Knowledge Conference}{March 03--07, 2025}{Dublin, Ireland}
\acmBooktitle{LAK25: The 15th International Learning Analytics and Knowledge Conference (LAK 2025), March 03--07, 2025, Dublin, Ireland}
\acmPrice{}
\acmDOI{10.1145/3706468.3706484}
\acmISBN{979-8-4007-0701-8/25/03}

\begin{document}

\title[Learner Control and Explainable Learning Analytics]{How Learner Control and Explainable Learning Analytics on Skill Mastery Shape Student Desires to Finish and Avoid Loss in Tutored Practice}

\author{Conrad Borchers}
\orcid{0000-0003-3437-8979}
\authornote{Both authors contributed equally to this research.}
\affiliation{
  \institution{Carnegie Mellon University}
  \streetaddress{5000 Forbes Ave}
  \city{Pittsburgh, PA 15213}
  \country{USA}
}
\email{cborcher@cs.cmu.edu}

\author{Jeroen Ooge}
\orcid{0000-0001-9820-7656}
\authornotemark[1]
\affiliation{
  \institution{Utrecht University}
  \streetaddress{Princetonplein 5}
  \city{Utrecht, 3584 CS}
  \country{The Netherlands}
}
\email{j.ooge@uu.nl}

\author{Cindy Peng}
\orcid{0009-0008-3599-2026}
\affiliation{
  \institution{Carnegie Mellon University}
  \streetaddress{5000 Forbes Ave}
  \city{Pittsburgh, PA 15213}
  \country{USA}
}
\email{cindypen@cs.cmu.edu}

\author{Vincent Aleven}
\orcid{0000-0002-1581-6657}
\affiliation{
  \institution{Carnegie Mellon University}
  \streetaddress{5000 Forbes Ave}
  \city{Pittsburgh, PA 15213}
  \country{USA}
}
\email{aleven@cs.cmu.edu}

\renewcommand{\shortauthors}{Borchers et al.}

\begin{abstract}
Personalized problem selection enhances student practice in tutoring systems. Prior research has focused on transparent problem selection that supports learner control but rarely engages learners in selecting practice materials. We explored how different levels of control (i.e., full AI control, shared control, and full learner control), combined with showing learning analytics on skill mastery and visual \textit{what-if} explanations, can support students in practice contexts requiring high degrees of self-regulation, such as homework. Semi-structured interviews with six middle school students revealed three key insights: (1)~participants highly valued learner control for an enhanced learning experience and better self-regulation, especially because most wanted to avoid losses in skill mastery; (2)~only seeing their skill mastery estimates often made participants base problem selection on their weaknesses; and (3)~\textit{what-if} explanations stimulated participants to focus more on their strengths and improve skills until they were mastered. These findings show how explainable learning analytics could shape students' selection strategies when they have control over what to practice. They suggest promising avenues for helping students learn to regulate their effort, motivation, and goals during practice with tutoring systems.
\end{abstract}

\begin{CCSXML}
<ccs2012>
   <concept>
       <concept_id>10003120.10003121</concept_id>
       <concept_desc>Human-centered computing~Human computer interaction (HCI)</concept_desc>
       <concept_significance>500</concept_significance>
       </concept>
   <concept>
       <concept_id>10010405.10010489</concept_id>
       <concept_desc>Applied computing~Education</concept_desc>
       <concept_significance>500</concept_significance>
       </concept>
 </ccs2012>
\end{CCSXML}

\ccsdesc[500]{Human-centered computing~Human computer interaction (HCI)}
\ccsdesc[500]{Applied computing~Education}

\keywords{explainable AI, mastery learning, intelligent tutoring systems, design-based research, K-12, self-regulated learning}

\maketitle

\section{Introduction}
Intelligent tutoring systems (ITS) improve learning and our understanding of learning processes~\cite{long2017enhancing,kulik2016effectiveness}. However, most ITS lack explainability and learner control because the algorithms powering them, such as Bayesian Knowledge Tracing, are not easily interpretable for learners~\cite{zhou2020assessing}. In response, research has offered learners more control over their learner models and learning materials~\cite{brusilovsky2023ai}. Open learner models allow learners to view and sometimes adjust personal data, such as inferred skill mastery, preferences, and progress~\cite{bull2020there}. Recent research on explainable AI (XAI) has revived the challenge of creating transparent systems~\cite{gunning2019darpa}. Various explanation types have been introduced to help people understand AI behavior~\cite{liao2020questioning}, including \textit{what-if} explanations which show how changing inputs or model parameters could affect outcomes~\cite{wexler2019whatif}.
 
Despite these advances, there is limited research on giving learners control over which learning materials ITS select~\cite{brusilovsky2023ai}. Furthermore, few ITS provide explainable learning analytics to help learners decide what, when, and how to practice. This gap is partly due to a lack of ITS research in out-of-class contexts like homework~\cite{azevedo2022lessons,aleven2004toward}, where students face unique challenges concerning self-regulated learning (SRL), including time management and motivation~\cite{xu2013students,bembenutty2011meaningful}. Therefore, we must understand how control over what to practice, combined with explainable learning analytics, could help students regulate their practice efforts. We designed an app that presents learning analytics about skill mastery to students and uses these analytics in \textit{what-if} explanations with learner control mechanisms. We evaluated our designs through semi-structured interviews with six middle school students, investigating two research questions:

\textbf{RQ1:} What are students' needs regarding learning analytics of skill mastery, \textit{what-if} explanations, and control over practice materials in the context of ITS practice?

\textbf{RQ2:} How do learning analytics of skill mastery and \textit{what-if} explanations shape students' selection strategies?

\section{Background and Related Work}

\subsection{Learning Analytics for Practice Support and Self-Regulated Learning}

Supporting learners in reflection and decision-making can foster self-regulated learning (SRL) and homework success~\cite{bembenutty2011meaningful}. Systems such as MetaTutor and Cognitive Tutor guide students in real-time, helping with goal-setting and metacognitive activities like help-seeking~\cite{azevedo2022lessons,aleven2004toward}. Adaptive scaffolds can also prompt learners to revisit and adjust plans~\cite{li2024analytics}. Despite these advances, applying SRL during homework poses challenges such as independently managing time, motivation, and environment~\cite{xu2013students}. Recent work has increasingly emphasized the role of contextual support beyond system-provided instruction during problem-solving. For example, the TeamSlides dashboard extends reflection beyond the immediate learning context by allowing students and teachers to revisit and analyze teamwork dynamics~\cite{echeverria2024teamslides}. However, this line of research has mostly focused on higher education~\cite{echeverria2024teamslides} and teacher behavior~\cite{borchers2024revealing}. Our work focuses on homework contexts and middle school students, which is relevant because SRL develops through homework~\cite{ramdass2011developing}.

\subsection{Learner Control in Tutoring Systems}

As ITS typically select problems based on student knowledge estimates, much research has focused on better estimating student knowledge and optimizing algorithms that use such estimates~\cite{huang2021general}. While effective in supporting mastery learning, this approach limits learner agency~\cite{deschenes2020recommender}, which could otherwise boost cognitive engagement~\cite{reeve2011agency}. To support decision-making about practice, research has explored giving learners more control over their practice activities by ``collaborating'' with AI~\cite{brusilovsky2023ai}. Open learner models showing students their mastery estimates are an influential example~\cite{bull2020there}, Studies show that learner control in interactive open learner models allowing for task selection can improve learning~\cite{long2017enhancing} and trust in educational systems, especially when the exercised control is visualized~\cite{ooge2023steering}. Therefore, our study builds on the premise that well-designed learner control can elevate learning. Research in educational psychology suggests shared control in task selection, with appropriate complexity, optimizes learning outcomes~\cite{corbalan2009combining}.

\subsection{Explainable AI in Education}

Explainable AI (XAI) is an increasing topic of interest in AI-supported education~\cite{khosravi2022explainable}. XAI aims to clarify the rationale behind algorithmic decision-making. Crucially, different audiences and contexts require different XAI solutions. For example, \citet{ooge2022explaining} showed that visual explanations for high school students can increase initial trust in an educational recommender system. Yet, \citet{szymanski2024feedback} found that teachers prefer relying on domain expertise over explanations when distractor-generating AI performs poorly. Providing another example of successful XAI, \citet{barria-pineda2021explainable} found that explanations increase student engagement with recommended learning content, leading to higher success rates. These advances suggest that XAI can help learners make informed decisions about their learning activities. XAI research with middle schoolers is rare. \citet{hitron2019can} suggest middle schoolers better grasp AI models by modifying inputs and observing related outputs. Accordingly, we design analytics enabling students to monitor mastery updates and choose problems to achieve expected mastery gains.

\section{Methods and Materials}

Our study investigates students' preferences for explainable learning analytics during AI-supported homework practice. We conducted semi-structured interviews with six middle school students to evaluate mock-up designs for an app that supports task selection during homework practice. These designs were based on prior research indicating that students prefer goal setting over other forms of homework support effective in offline homework settings~\cite{peng2024designing}. The study was approved by the local institutional review board and included informed consent from participants and their guardians.

\subsection{Math Practice App}

\begin{figure*}[htp]
	\centering
	\includegraphics[width=0.85\linewidth]{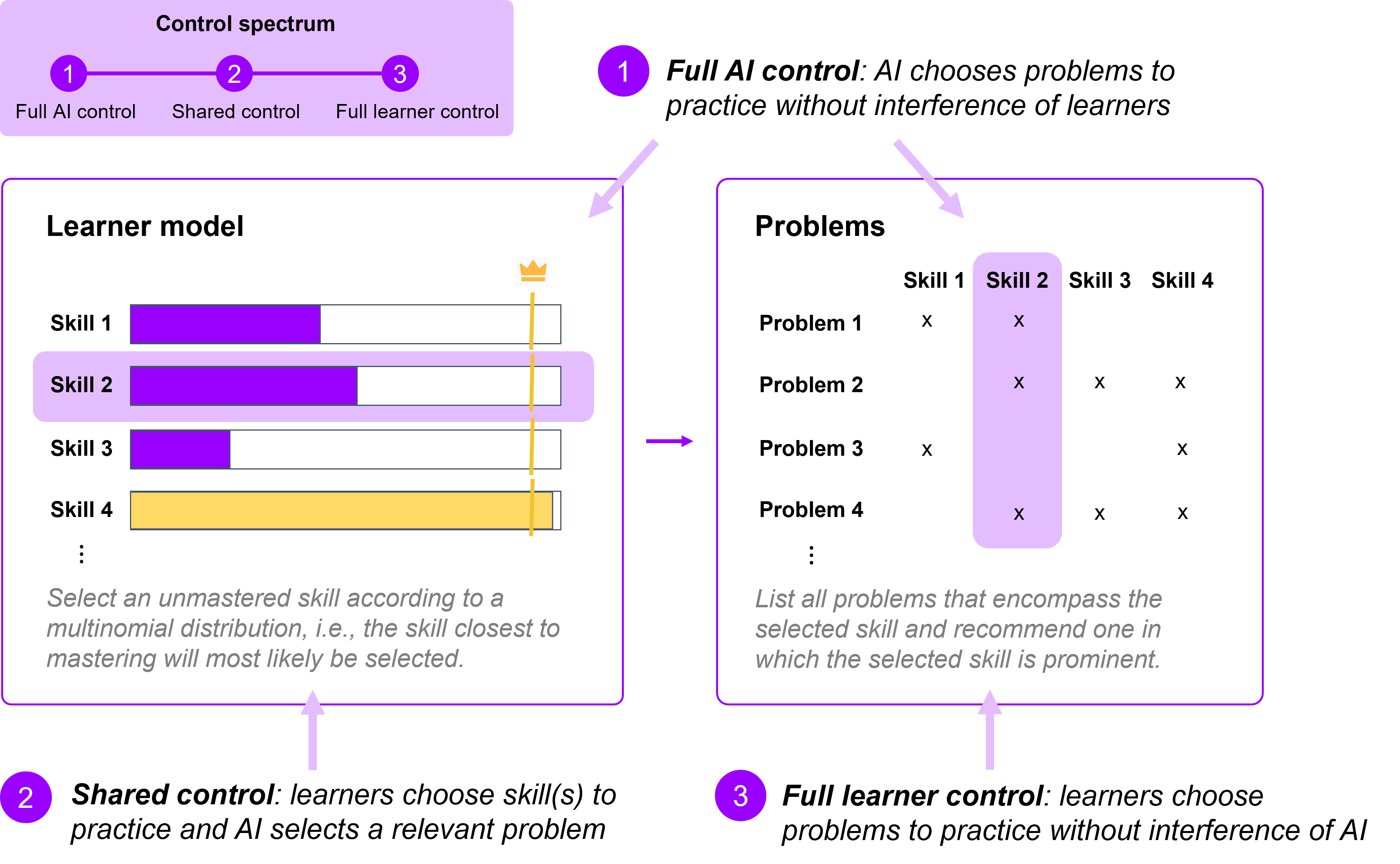}
	\caption{The recommendation algorithm underlying our app is based on skills associated with math problems. We designed three granularity levels of control: full AI control, shared control, and full learner control.}
	\label{fig:controlspectrum}
\end{figure*}

We designed a mock-up app for linear equation practice, building on Lynnette~\cite{long2017enhancing}. Like other ITS, Lynnette represents student knowledge as skills required to solve multi-step problems~\cite{koedinger2016testing} and uses Bayesian Knowledge Tracing to estimate skill mastery~\cite{huang2021general}. Probabilities express if learners know specific skills, with 0.95 considered mastery. The app recommends problems with skills close to mastery (\Cref{fig:controlspectrum}) and has three key features (\Cref{fig:results}).

\textit{Control.} We designed three \feature{learner control} mechanisms, exploring how decision-making can be distributed between AI and learners, as shown in \Cref{fig:controlspectrum}. \textit{Full AI control} means the algorithm determines the next problems, and learners cannot intervene. \textit{Shared control} allows learners to choose the skills they want to practice, but the algorithm then picks a problem related to those skills. \textit{Full learner control} means learners can directly choose the next problem.

\begin{figure*}[htp]
	\centering
	\includegraphics[width=0.85\linewidth]{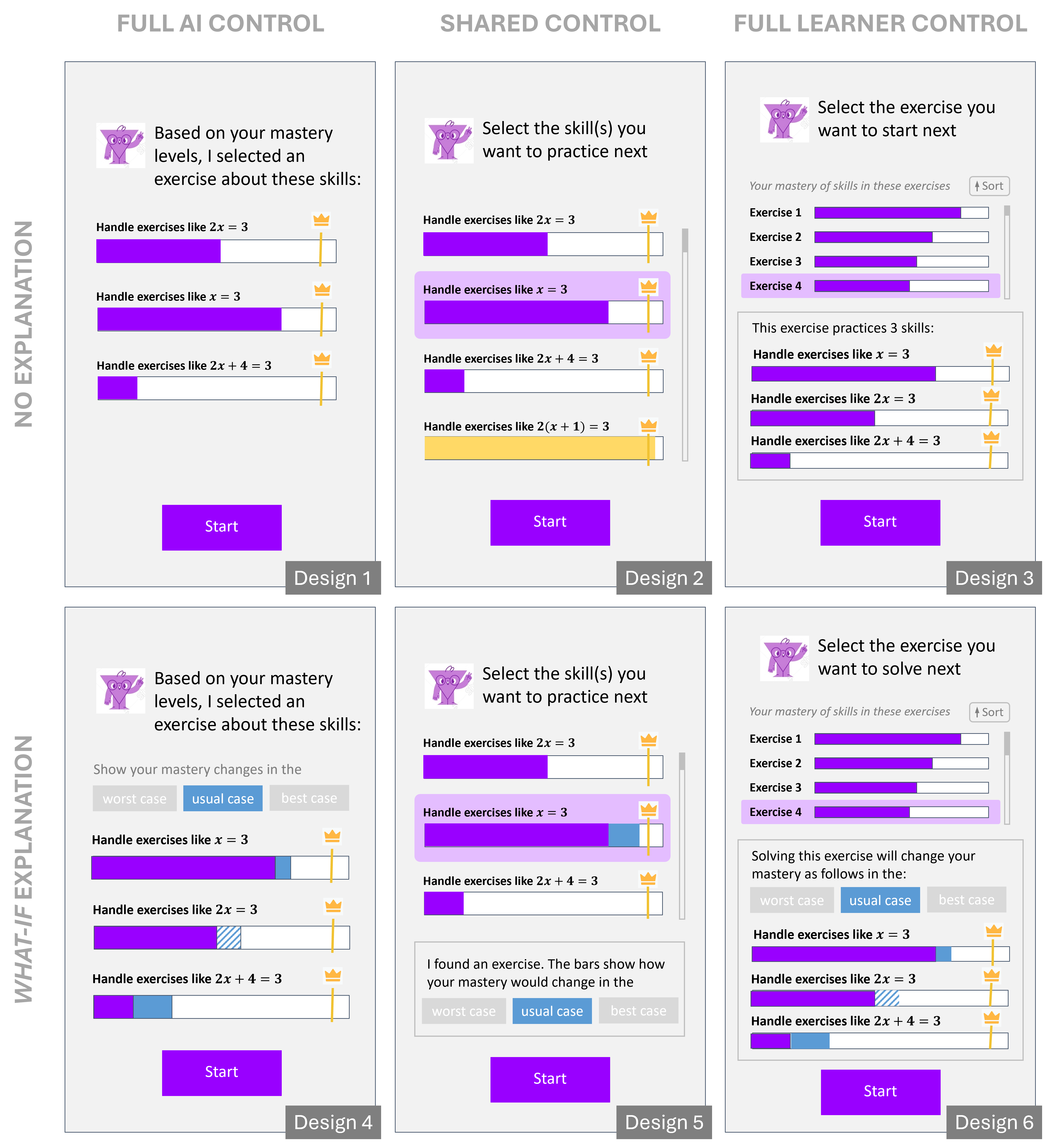}
	\caption{Our six designs provide one of three control mechanisms and visual \textit{what-if} explanations.}
	\label{fig:designs}
\end{figure*}

\textit{Transparency.} \Cref{fig:designs} illustrates how our designs aim to combine \feature{transparent skill mastery} with \feature{\textit{what-if} explanations} to foster transparency. First, bar charts visualize learners' skill mastery, with gold bars indicating mastered skills. In the interfaces with full learner control, mastery estimates are averaged for all skills involved in a problem. All bars were animated: smooth increases and decreases helped participants understand the changes. Second, we adopted the idea of \textit{what-if} explanations, which give insight into what an AI system would predict if a given input would change~\cite{liao2020questioning,wexler2019whatif}. In our tutoring system, this corresponded to updated mastery estimates given the correctness of problem steps. Explanations were split into the worst and best case (i.e., completing all problem steps incorrectly or correctly) and the usual case (i.e., the expected change based on past performance data).

\subsection{Participant Recruitment, Procedure, and Data Analysis}

We recruited six middle school students through parent outreach centers in the Northeastern USA. \Cref{fig:results} contains their demographics. The sample size, comparable to other educational design research~\cite{nguyen2024designing}, was determined based on project resources and data saturation, meaning sampling stopped when few new themes emerged during interviews. During one-hour online semi-structured interviews between July and September 2023, participants first followed a tutorial that introduced the bars visualizing their skill mastery and changes thereof. Then, they interacted with all our six designs in the same order as shown in \Cref{fig:designs} while answering questions about usefulness, their needs, and understanding of the components (e.g., ``Do you like selecting exercises yourself here? Why or why not?''). All materials are available online.\footnote{\url{https://github.com/conradborchers/mastery-xai-lak25}} The interviewers did not intervene when participants made inaccurate interpretations.

Following Braun and Clarke~\cite{braun2018thematic}, two researchers conducted a thematic analysis using the Dovetail software. They first independently coded the interview transcripts, combining an inductive approach with a deductive approach focused on participants' understanding of our designs, selection strategies, and needs regarding explanations and control. Next, they iteratively consolidated the codes into themes through independent affinity diagramming and team discussions.

\section{Results}

No major usability issues arose during the interviews, apart from some initial confusion about the meaning of the cases in the \textit{what-if} explanations. For example, P1 and P4 presumed, \quote{Worst case is, these are the topics you're horrible at, and best case is, these are the ones you're good at} (P4). However, after interacting with the case buttons, all participants seemed to understand the cases correctly. Only P2 developed an alternative interpretation: \quote{[The cases] show \rem the worst you have ever done on the exercise, your usual performance on the exercise, and your best.}

\begin{figure*}
    \centering
    \includegraphics[width=\linewidth]{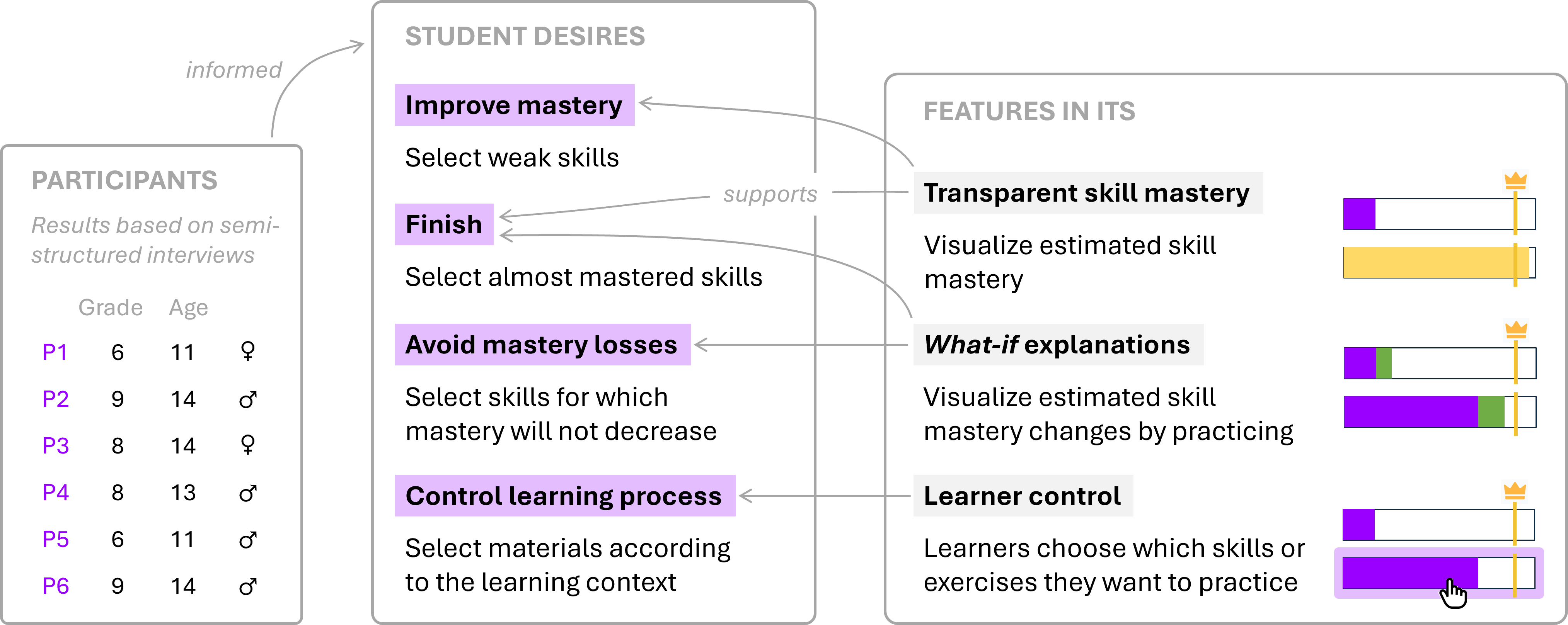}
    \caption{Summary of the four main desires participants expressed during the interviews and how our system supports them.}
    \label{fig:results}
\end{figure*}

Our thematic analysis yielded four main student needs (RQ1), which shaped problem selection strategies (RQ2): the desire to improve mastery, the desire to finish, the desire to avoid mastery losses, and the desire to control the learning process. \Cref{fig:results} shows how our ITS features support these desires, and the following paragraphs elaborate on each.

\begin{shaded}\noindent
\textbf{Desire to improve mastery:} Participants tended to diagnose weaknesses when observing analytics of skill mastery and selected skills and problems to compensate for weaknesses.
\end{shaded}

Participants often mentioned the \feature{transparent skill mastery} estimates to guide their efforts and check whether skills required attention. For example, P3 described a cyclic process of (re)visiting mastery estimates: \quote{I would want to see the skills first, just so I could see what I need to practice. After that, I'd want to do the practice and then maybe see the skills again, how they changed afterwards.} This indicates participants perceived mastery estimates not merely as passive feedback but as a diagnostic tool for practice and gauging progress over time.

Moreover, participants often identified weaknesses they \quote{had to work on} when seeing skill mastery estimates. For example, P5 described using mastery estimates to \quote{know what to work on \rem I'll just work on that to get that better because that's not best. But I wouldn't work on [a skill...] if I already mastered all of that stuff; then you don't need to check it.} Similarly, other participants mentioned: \quote{If I wanted to pick a skill to practice next, I'd probably do a thing that I wasn't the best at} (P2) and \quote{Oh, I have to work on this to get it up} (P3). Thus, low mastery estimates could evoke a desire to improve. However, P2 alluded to the idea that indicating low mastery might also discourage selecting challenging problems: \quote{If it's the world's hardest skills in there, you might not want to show that.}

\begin{shaded}\noindent
\textbf{Desire to finish:} Participants felt incentivized to practice almost-mastered skills to quickly achieve mastery or because they anticipated low effort to practice such skills.
\end{shaded}

Participants used the \feature{transparent skill mastery} estimates to detect which skills were close to completion and seek opportunities to quickly achieve mastery. P2 remarked: \quote{If you're close to mastering it, that could give you some more incentive to do it.} Similarly, P3 shared: \quote{I would go with this middle one, just because I am about to master it.} This suggests visual progress cues motivated participants to complete tasks, especially when mastery was within reach. \feature{\textit{What-if} explanations} reinforced this effect as they previewed how much progress learners could make with the next problem. P2 mentioned: \quote{I'm immediately drawn to the one [skill] I'm doing the best at \rem because you can choose the best-case scenario, and you end up just short of mastering it.} This reflects how participants used the \textit{what-if} insight for future mastery estimates to seek attainable goals, often selecting skills where success seemed within reach.

Additionally, some participants favored skills they had already mastered, as related problems felt easier. For instance, P1 explained: \quote{The [skill] I've mastered the most would be the easiest.} This suggests participants were motivated by the perceived low effort required to complete problems involving highly mastered skills.

\begin{shaded}\noindent
\textbf{Desire to avoid mastery losses:} Participants frequently indicated they wanted to keep mastery increasing and used \textit{what-if} explanations extensively for decision-making.
\end{shaded}

Four participants often based their decision-making on mastery loss aversion. For example, P4 explained: \quote{If [mastery] went down, we would try to work on it, and if it went up, then we would probably work on it just a teeny bit, but not as much as if it went down.} This suggests declining skill bars can function as triggers for corrective action through task selection, and loss aversion could be a stronger motivator than progress. Participants linked this loss aversion to appreciating \feature{\textit{what-if} explanations}, which helped them to \quote{know what you're going to dive into and how the exercise is going to help you} (P4). For example, P6 said: \quote{I would find [\emph{what-if} explanations] useful a lot because I'm always worried that when I do an assignment \rem I'm gonna go down to 0 or something if I do bad.} Similarly, P1 explained why they liked seeing in advance what would happen with their skill mastery: \quote{I would try to avoid losing a lot of the percentage in [the skill bars], so I think it would be helpful to show \rem the worst-case scenario for the ones that have the lower score already. I feel if I would try that one, \rem it could go down a lot, which I would not want to happen.} In the case of a predicted mastery decrease, P2 even expected to \quote{never select that [skill] again} because it \quote{shows you suck at something.}

Interestingly, all participants marked the usual case as their favorite case in the \textit{what-if} explanations, which shows \quote{the best prediction of what will happen when you complete this} (P2). Others seconded this: \quote{I feel the usual case is the most helpful because it's what I usually am doing the practices like} (P3) and \quote{It tells you, this is on average what you do} (P4). Two participants said they would use the usual case for decision-making: \quote{I would pick the exercise [or skill] where the usual case progresses me the most \rem because I don't have to push myself too much, \rem just a teeny bit each day} (P4), and \quote{If it believes that I would progress quite a bit, I would probably choose it \rem because it would be more helpful if I want to finish something} (P6). Finally, P4 added that they also appreciated the best case as \quote{it's telling me this is the best you can do on a daily [basis]} and P1 suggested an interesting use-case for the worst case: \quote{On my bad math days I'd probably pick the worst case and the usual case, so I could know what would happen if I was not doing as well as on usual days.} This indicates P1 managed their expectations for mastery growth based on performance scenarios.

\begin{shaded}\noindent
\textbf{Desire to control the learning process:} Participants valued sharing control with AI over their learning process so they could select materials according to their learning context.
\end{shaded}

Participants strongly preferred designs with \feature{learner control} because it enhanced their learning experience and self-regulation. For example, P6 mentioned that \quote{understanding and choosing skills \rem helps me learn more and study better versus everything being chosen for me, and kind of being forced to do one, regardless of how I feel about it,} while P3 appreciated control \quote{because then I can pick what kind of math problems I would like to do for my mood.} Furthermore, P2 asserted: \quote{You should have some level of control over what you want to do in the app.} 

Students had diverse reasons for preferring control. First, participants preferred adjusting difficulty when their motivation to practice was low or when they struggled. P3 mentioned: \quote{Probably when I'm feeling lazy and more like I just wanna do something that is easier, I'd probably pick the higher bars.} Similarly, P5 justified a preference for easier skills by stating: \quote{It depends on if I am more tired or have more things to do that day.} P1 explicitly preferred being in control to regulate difficulty depending on their mood and state: \quote{[On my best math days] I would just let them pick a skill for me, and then on my bad math days, I would pick a skill that would be easier for me to solve.}  Second, several participants explained how they would prioritize practicing skills based on what their teacher expected: \quote{It completely depends if it's just supposed to finish one exercise, or you're trying to get all exercises to a specific point} (P6) and \quote{The ones assigned I would prioritize, like the homework. Or what's gonna be on upcoming tests or something like that} (P3). 

The preference for control came with four reservations. First, Design 6 can be overwhelming: \quote{It does feel a little bit too much \rem I don't think you really need [the bars shown next to exercises]} (P6). Thus, adding analytics that support decision-making might be unnecessary if \textit{what-if} explanations are already present. Second, P1, P3, and P5 appreciated recommendations that \quote{based on your levels, are showing what's best for you to learn} (P1) because they would not always know what to select. Third, P2 and P4 realized they could make suboptimal choices or need help prioritizing skills: \quote{The tutor [should] force you to do something good \rem when you pick easy additions like $1+1$\!} (P2). Finally, P6 found it \quote{more important to have choices [i.e., control] than just having the [\textit{what-if} explanations].}

\section{Discussion}

Our study investigated how explainable learning analytics~\cite{li2024trustworthy} can support students' practice decisions during tutored mathematics homework, where SRL is key for learning outcomes~\cite{peng2024designing,cadime2017homework}. We designed such analytics based on skill mastery estimates and interviewed six middle school students about their needs and selection strategies during practice.

\subsection{Students Desire Practice Control, Mainly To Avoid Mastery Losses}

RQ1 centered around students' needs regarding explainable learning analytics based on current and future skill mastery during homework. Students expressed a strong desire for control over what they practice. The reasons were manifold, including the ability to adjust difficulty based on their motivation and mood, the need to align practice with teacher expectations or upcoming assessments, and the preference to avoid tasks that felt too challenging when their motivation was low. Many students were particularly concerned about a decline in their mastery levels and were motivated to take corrective actions to prevent such a decline. This finding extends earlier research on open learner models, which indicates that students feel rewarded by improving their mastery estimates in ITS~\cite{long2011students}. Importantly, given that recent higher education research has found that students may disengage to avoid failing at difficult coursework~\cite{shabab2024understanding}, students with loss aversion may disengage in practice if analytics predict they may not succeed in a task. Therefore, as our findings highlight, it is important to offer such students alternative practice tasks that allow them to gain mastery and succeed. Indeed, students in our sample frequently used the \textit{what-if} explanations to anticipate potential decreases in mastery and made strategic choices to increase their current mastery levels by choosing different tasks with better-expected outcomes. This finding, together with the work of \citet{shabab2024understanding}, implies that explainable learning analytics and reward prospects might increase student engagement with ITS during homework practice. At the same time, students primarily mentioned preferring selecting skills, which implies the system would select specific problems. This aligns with past research noting that learners can be cognitively overloaded by large amounts of choice~\cite{corbalan2009combining}. Moreover, past work on open learner models in ITS showed that students' learning benefits from feedback on mastery changes~\cite{long2016mastery}. Our research adds that motivation might be improved by visualizing changes in mastery in shared control settings. 

\subsection{Explainable Learning Analytics Can Spur Desires to Improve and Finish Practice}

RQ2 centered around how explainable learning analytics shape students' practice selection strategies. Our results reveal a contrast between traditional skill mastery analytics and \textit{what-if} explanations. On the one hand, when students only saw skill mastery analytics, their focus often gravitated toward areas of weakness. Past work observed that such a focus is especially pronounced in students with low self-efficacy in mathematics~\cite{abdulla2024should}, making it important to create experiences of success for these students. On the other hand, introducing \textit{what-if} explanations seemed to transform participants' decision-making, shifting their focus to skills closer to mastery so they could achieve mastery and finish their practice. This is desirable as practicing problems that focus on easier skills (i.e., skills close to mastery) is expected to lead to faster mastery attainment~\cite{huang2021general}. Furthermore, focusing on skill mastery and achievable progress, rather than weaknesses, aligns with growth mindset principles that encourage persistence and confidence~\cite{hochanadel2015fixed}. Our results suggest that \textit{what-if} explanations can reinforce these principles by helping students visualize progress, boosting motivation, and shifting their focus toward achievable goals. Growth mindset interventions enhance academic achievement according to a recent meta-analysis~\cite{burnette2023systematic}, making this an encouraging direction for future research.

Notably, occasional challenges in understanding \textit{what-if} explanations suggest that additional support may be needed to help learners grasp the concept of alternative outcomes. This is especially relevant in middle school settings, as the abstract thinking required to envision alternative realities develops throughout adolescence~\cite{beck2006children}. Further research is needed to explore how design can address potential misconceptions, for example, by explicitly conveying where students currently stand and where they could stand in the future through longitudinal graphs with practice dates~\cite{nagashima2022designing}. Additionally, further design considerations are needed to ensure learners are not demotivated by mastery decreases.

\subsection{Limitations and Future work}

Our study has several limitations that restrict the generalizability of our results. First, all participants were white, and only one was neurodiverse. A more diverse sample could uncover additional student needs and different selection strategies for learning materials. Future work may also study how high school students perceive our proposed analytics, as they may exhibit different SRL strategies during homework compared to middle school students~\cite{hong2009homework}. Second, our interviews only captured students' expectations, but what students say can differ from what they would do in real-life learning settings. Thus, larger-scale experiments involving implemented versions of our designs should monitor how students actually use \textit{what-if} explanations and control mechanisms and study whether they benefit in terms of learning, motivation, and self-regulation. Third, we have not consulted teachers and other pedagogical experts whose homework expectations may be incompatible with student preferences expressed in this study. Future work may design for practice goal selection where teachers and parents can be involved in students' practice decisions. Fourth, we conceptualized shared learner control such that combining it with \textit{what-if} explanations was natural. Future studies could investigate how other forms of learner control (e.g., postponing problems or rejecting recommendations) can be combined with different types of explanations and how those combinations affect student attitudes and practice selection strategies. Finally, our skill bar designs did not visualize how often learners had practiced corresponding skills. As a result, learners could not differentiate between low skill estimates due to insufficient practice or due to true weak skill mastery after repeated practice. Designing toward clarifying that distinction merits future research based on our findings.

\section{Conclusions}

This study examined how explainable learning analytics influence middle school students' practice needs and strategies in intelligent tutoring systems. Participants strongly desired control over their learning process and actively used mastery estimates and \textit{what-if} explanations to inform their decisions. They used these analytics to identify weaknesses, focused on skills nearing mastery, and adjusted their practice based on their motivational states. While students desired control, results suggest they find it sufficient to select skills and let the system select problems, as opposed to selecting problems themselves. Importantly, \textit{what-if} explanations reinforced participants' efforts to attain mastery instead of being driven by avoiding mastery losses. Our findings suggest that explainable learning analytics could help enhance SRL by aligning with students' desires to improve their skill mastery and maintain control over their learning process. Our results also merit further exploration into how to best visualize alternative practice outcomes for middle school students through explainable learning analytics.

\begin{acks}
This research was funded by the Institute of Education Sciences (IES) of the U.S. Department of Education (award \#R305A220386) and was initiated during a research visit funded by Research Foundation Flanders (grant V431323N).
\end{acks}

\bibliographystyle{ACM-Reference-Format}
\bibliography{main}

\end{document}